\documentstyle[preprint,aps]{revtex}
\begin{document}
\draft
\title{Enhancement of Optical Nonlinearity Through \\
Anisotropic Microstructures}
\author{K. P. Yuen$^1$, M. F. Law$^1$,
K. W. Yu$^{1}$\footnote{Corresponding author. e-mail: kwyu@phy.cuhk.edu.hk},
and Ping Sheng$^2$}
\address{$^1$Department of Physics, The Chinese University of Hong Kong, \\
         Shatin, New Territories, Hong Kong}
\address{$^2$Department of Physics, The Hong Kong University of Science and
         Technology, \\ Clear Water Bay, Kowloon, Hong Kong}
\maketitle

\begin{abstract}
We investigate the polarization dependence of optical nonlinearity
enhancement for a uniaxial anisotropic composite of metal nanocrystals
in a dielectric host.
Three cases are distinguished depending on whether the polarization is
parallel, perpendicular or unpolarized with respect to the axis of
anisotropy.
For the parallel polarization, the results show that the 3D results are
qualitatively similar to the 2D case reported recently.
For the perpendicular polarization, the results are markedly different
from the parallel counterpart: In contrast to the absorption, the
enhancement factor actually increases with the anisotropy. Thus
the separation of the absorption and enhancement peaks becomes even more
pronounced than the parallel polarization case.
These results indicate a strong polarization dependence of the nonlinear
optical response.
\end{abstract}
\vskip 5mm
\pacs{PACS Numbers: 82.70.Dd, 72.20.Ht, 42.65.$-$k, 61.90.+d}

\section{Introduction}

With the advent of high-power coherent light sources, optical nonlinearity
has become a rapidly growing field both in basic research as well as in
practical applications \cite{JOSA}. The usefulness of optical nonlinearity
depends on the availability of materials with a large nonlinear
susceptibility and a short response time. While the search for a homogeneous
material that possesses a large nonlinear susceptibility continues, the use
of composite materials has been proposed as an alternative for enhancing the
optical nonlinearity, through the inhomogeneity of the local electric field
and/or the surface-plasmon resonances \cite{JOSA}.
These enhancement mechanisms are sensitive to the composite microstructure.

In a recent work \cite{Yuen}, we proposed the use of structurally
anisotropic composites to enhance the optical nonlinearity.
Anisotropy is clearly a phenomenon common in most materials, and in optical
phenomena. Anisotropy can be an intrinsic material property or can be
induced, e.g., by the application of fields.
The nonlinearity can be enhanced by using nonspherical (e.g., ellipsoidal)
particles \cite{Haus}.
Induced nonlinearity such as the electrostrictive mechanism (through
gradients in EM fields) that operates in the microsecond regime was recently
reviewed \cite{Neeves}.

In Ref.\cite{Yuen}, we considered composites made of components which are
locally isotropic in their material properties, but with electric-field
induced anisotropic microstructures.
Here the characteristic length scale of the microstructure is assumed to
be much less than the wavelength, so that the optical response of the
system is characterized by an effective dielectric tensor.
We showed that by controlling the magnitude of the applied electric field,
for example,
one can vary the degree of geometric anisotropy and thereby tune the system
to achieve maximum optical nonlinearity in accordance with the properties
and volume fraction of nonlinear materials \cite{Yuen}.
Furthermore, we proposed the use of the electrorheological (ER) effect
\cite{ERF,PRL} to realize the desired geometric anisotropy, e.g.,
during the annealing process, in material systems containing a nonlinear
optical component.
An important benefit of the anisotropic microstructures considered in this
work and Ref.\cite{Yuen} is the separation of the nonlinear $\chi^{(3)}$
enhancement peak from the absorption peak, thus raising the application
potential of the microstructure enhancement mechanism.

While preliminary results were presented for the parallel polarization in
2D \cite{Yuen}, it is also necessary to obtain the results for the
perpendicular polarizations and hence results for the unpolarized response.
Moreover, it is also instructive to examine the more realistic 3D case.
In what follows, we shall use the spectral representation of local fields
\cite{spectral} to model the anisotropy enhancement effect.
The spectral representation offers the advantage of separation of material
parameters from the microstructure information, contained in the spectral
density function $m(s)$.
It has been shown recently that the optical nonlinearity enhancement effect
is directly linked to the behavior of $m(s)$ \cite{Sheng}.

The paper is organized as follows.
In the next section, we present the spectral representation of anisotropic
composite media.
In section III, we present results for the absorption and enhancement peaks
in the optical response, followed by a summary of our results.

\section{Spectral Representation of Anisotropic Composite Media}

We consider a composite metal-dielectric system, with metal being the
nonlinear optical component.
We assume the particle size is much smaller than the wavelength of light
so that the quasi-static approximation can be used.
The local constitutive relation is given by
${\bf D}=(\epsilon+\chi |{\bf E}|^2) {\bf E}$,
where $\epsilon$ is the (position dependent) dielectric constant and $\chi$
is the (position dependent) third-order Kerr nonlinear susceptibility.
We model this material system by a lattice model, with assigned probability
of metal particle occupation at each lattice site. To introduce uniaxial
anisotropy, we assign two different occupation probabilities parallel
($p_\parallel$) and perpendicular ($p_\perp$) to a particular direction of
an occupied site.
We choose several values of the anisotropy ratio $r=p_\perp/p_\parallel$,
varying from isotropic ($r=1$) to highly anisotropic ($r=0$).
We invoke the effective-medium approximation (EMA) for anisotropic media
\cite{Bernasconi,Berthier}.
The coupled EMA self-consistency equations read \cite{Bernasconi,Berthier}:
\begin{eqnarray}
p_\parallel{\epsilon_1-\epsilon_\parallel \over
            \epsilon_1+z_\parallel \epsilon_\parallel} +
(1-p_\parallel){\epsilon_2-\epsilon_\parallel \over
            \epsilon_2+z_\parallel \epsilon_\parallel} = 0, \label{EMA1} \\
p_\perp{\epsilon_1-\epsilon_\perp \over
            \epsilon_1+z_\perp \epsilon_\perp} +
(1-p_\perp){\epsilon_2-\epsilon_\perp \over
            \epsilon_2+z_\perp \epsilon_\perp} = 0, \label{EMA2}
\end{eqnarray}
where $z_\parallel$ and $z_\perp$ are the parameters parallel and
perpendicular to the applied dc field.
In Eqs.(\ref{EMA1}) and (\ref{EMA2}),
$\epsilon_1$ denotes the dielectric constant of metal, and $\epsilon_2$ that
of the dielectric component; $\epsilon_\parallel$ and $\epsilon_\perp$
denote, respectively, the parallel and perpendicular components of the
effective dielectric tensor.
In 2D, the $z$-parameters are given by \cite{Bernasconi}
\begin{eqnarray}
z_\parallel={\tan^{-1} \sqrt{\epsilon_\perp/\epsilon_\parallel} \over
             \tan^{-1} \sqrt{\epsilon_\parallel/\epsilon_\perp}},
             \label{z2D1} \\
z_\perp={\tan^{-1} \sqrt{\epsilon_\parallel/\epsilon_\perp} \over
             \tan^{-1} \sqrt{\epsilon_\perp/\epsilon_\parallel}}.
             \label{z2D2}
\end{eqnarray}
Let $U=\epsilon_\perp/\epsilon_\parallel$.
In 3D, the $z$-parameters are given by \cite{Bernasconi}
\begin{eqnarray}
z_\parallel={\tan^{-1} \sqrt{2U+U^2} \over
             \tan^{-1} 1/\sqrt{2U+U^2}}, \label{z3D1} \\
z_\perp={\tan^{-1} \sqrt{2U+U^2}/U \over
             \tan^{-1} U/\sqrt{2U+U^2}}. \label{z3D2}
\end{eqnarray}
The above theory is applicable to symmetric microstructure as in granular
composites, and is an extension of the Bruggeman theory to anisotropic
composite media \cite{Berthier}.
The depolarization factors are related to the $z$-parameters through
$D_\parallel=1/(1+z_\parallel)$ and $D_\perp=1/(1+z_\perp)$.
There is a simple relation for the depolarization factors for uniaxial
anisotropy in $d$ dimensions \cite{Berthier}:
$$
D_\parallel + (d-1) D_\perp = 1,
$$
which can be readily verified by using Eqs.(\ref{z2D1})--(\ref{z3D2}).
For the dispersion microstructure, as in colloidal suspensions, it is more
appropriate to use the Maxwell-Garnett approximation \cite{Berthier}:
\begin{eqnarray}
{\epsilon_\parallel-\epsilon_2 \over
            \epsilon_\parallel+z_\parallel \epsilon_2}
=p_\parallel{\epsilon_1-\epsilon_2 \over
            \epsilon_1+z_\parallel \epsilon_2}, \\
{\epsilon_\perp-\epsilon_2 \over
            \epsilon_\perp+z_\perp \epsilon_2}
=p_\perp{\epsilon_1-\epsilon_2 \over
            \epsilon_1+z_\perp \epsilon_2}.
\end{eqnarray}
Below we consider only the EMA case.
For two-component composites, it has proved convenient to adopt the spectral
representation of the effective linear response \cite{spectral}:
Let $v=1-\epsilon_1/\epsilon_2$, $w_\alpha=1-\epsilon_\alpha/\epsilon_2$
($\alpha$ = $\parallel$ or $\perp$), and $s=1/v$, we find
\begin{eqnarray}
w_\alpha(s)=\int_0^1 {m_\alpha(s'){\rm d}s' \over s - s'},
\label{sp}
\end{eqnarray}
where $m_\alpha(s')$ is the spectral density which is obtained through a
limiting process:
\begin{eqnarray}
m_\alpha(s') = \lim_{\eta \to 0^+} -{1\over \pi}{\rm Im}\
w_\alpha(s'+i\eta).
\end{eqnarray}
In what follows, we consider the 3D case.
The EMA can readily be solved in the spectral representation.
In our numerical calculations, we choose the real part at several hundred
equally spaced values across the interval $0\le s'\le 1$, and the imaginary
part $\eta$ to be some small positive value. The actual value of $\eta$ is
unimportant. We found that $\eta=0.001$ gives acceptable results by checking
the sum rule:
\begin{eqnarray}
\int_0^1 m_\alpha(s'){\rm d}s' = p_\alpha.
\end{eqnarray}
From any given value of $s$, the ratio $\epsilon_1/\epsilon_2$ is calculated
from $s=s'+i\eta$, and subsequently $\epsilon_\alpha$, hence $w_\alpha$, may
be evaluated numerically.
In what follows, we denote $p_\parallel=p$ for convenience.

For isotropic composites, $r=1$, $z_\parallel=z_\perp=2$ and the EMA
self-consistency equations can be solved analytically.
The spectral density is given by \cite{Sheng}:
\begin{eqnarray}
m(s',p)={1\over 2}(3p-1) \theta (3p-1) \delta (s') +
{3\over 4\pi s'}\sqrt{(s'-s_1)(s_2-s')},
\end{eqnarray}
valid for $s_1 < s' < s_2$, where
\begin{eqnarray}
s_1={1\over 3}\left( 1+p-2\sqrt{2p(1-p)}\right), \ \ \
s_2={1\over 3}\left( 1+p+2\sqrt{2p(1-p)}\right).
\end{eqnarray}
We first consider the parallel response, $\epsilon_\parallel$.
In Fig.\ref{fig1},
we plot the spectral density $m_\parallel(s)$ (associated with
$\epsilon_\parallel$) against $s$ for several values of $p$ and $r$.
For small $p$, e.g., $p=0.1$, the metal nanoparticles form essentially
isolated clusters and the spectral function exhibits a single peak centered
around $s=0.2$. More interestingly, as the anisotropy increases, the
spectral peak becomes narrower. This can be understood by the fact that
when $r \to 0$, more chain-like isolated clusters are formed, leading to
a narrow spectral peak.

When $p$ gradually increases, the interaction among clusters occur, leading
to a broad spectral peak in the isotropic case. However, the spectral peak
remains relatively narrow in the anisotropic cases. When $p$ increases
further, the metal clusters begin to percolate the system in the isotropic
case, leading to a delta-function contribution to the spectral density at
$s=0$. As $p>1/3$, we observe the familiar edge singularity
near $s=0$ for the isotropic case.
As is evident from the figures, the 3D results are qualitatively similar to
the 2D ones \cite{Yuen} except for some fine details.
From the results, it is clear that anisotropy can change the spectral
function substantially. This is the source of the pronounced effect on
nonlinearity enhancement.

We next present the perpendicular response, $\epsilon_\perp$.
In Fig.\ref{fig2},
we plot the spectral density $m_\perp(s)$ (associated with $\epsilon_\perp$)
against $s$ for several values of $p$ and $r$ as in Fig.1.
For small $p$, e.g., $p=0.1$, again the metal nanoparticles form essentially
isolated clusters. The spectral density is generally smaller in magnitude.
More interestingly, as the anisotropy increases ($r \to 0$), $m_\perp(s)$
decreases with $r$. This is due to the fact that chain-like clusters are
isolated from one another in the perpendicular direction.
As $r$ decreases, the spectral function splits to a two-peak structure.
For larger $p$, the decrease of $m_\perp(s)$ with $r$ persists.
Since the absorption is directly proportional to $m_\perp(s)$, we obtain an
interesting result that the absorption actually decreases with the strength
of anisotropy.

\section{Polarization Dependences of the Absorption and the $\chi^{(3)}$
Enhancement Peak}

When the spectral density $m_\alpha(s)$ is known as a function of $s$ and
$p$, the effective linear response can be calculated from Eq.(\ref{sp}).
We adopt the Drude model for the dielectric function of metal nanoparticles:
\begin{eqnarray}
\epsilon_1(\omega)=1-{\omega_p^2 \over \omega(\omega+i\gamma)},
\end{eqnarray}
where $\omega_p$ is the plasma frequency and $\gamma$ is the damping
constant. We choose $\gamma=0.01\omega_p$ and $\epsilon_2=1.77$ for our
model calculation. In this work, we assume only the metallic component to
be nonlinear.

In Fig.\ref{fig3},
the absorption peak Im($\epsilon_\parallel$) is plotted against frequency
$\omega$ for various $r$ and $p$.
In Fig.\ref{fig4},
the absorption peak Im($\epsilon_\perp$) is plotted against frequency
$\omega$ for various $r$ and $p$.
As evident from the results, the absorption peak exhibits similar behavior
as the spectral density does. This is attributed to the fact that the
absorption is related to the imaginary part of the effective dielectric
function. From the results, it is also clear that anisotropy can have an
important effect on absorption. We next study the effect of anisotropy on
the $\chi^{(3)}$ enhancement factor.

If a plane-polarized electromagnetic wave of amplitude $E_0$ with the
polarization parallel the uniaxial anisotropy axis is incident upon the
composite system, the local field averages are given by \cite{Sheng}
\begin{eqnarray}
p \langle E_1^2\rangle  &=&\int_0^1 {\rm d}s'{s^2 m_\parallel(s')
\over (s - s')^2} E_0^2, \\
p \langle |E_1|^2\rangle  &=&\int_0^1 {\rm d}s'{|s|^2 m_\parallel(s')
\over |s - s'|^2} E_0^2.
\end{eqnarray}
From the average local fields, we calculate the effective nonlinear response
as:
\begin{eqnarray}
\chi_\parallel |E_0|^2 E_0^2 = p \chi_1 \langle |E_1|^2\rangle \langle
E_1^2\rangle.
\end{eqnarray}
This expression results from the mean-field approximation \cite{Sheng}.

In Fig.\ref{fig5},
we plot the enhancement factor, $|\chi_\parallel|/\chi_1$, against $\omega$
for various $r$ and $p$.
Here we observe that anisotropy has indeed a pronounced effect on the
enhancement peak, as expected. As $p$ increases, the enhancement peak
exhibits a red shift, in analogy to the Maxwell-Garnett microstructure.
These results are qualitatively similar to the 2D ones reported recently
\cite{Yuen}.

Similar considerations apply to the perpendicular polarization.
In Fig.\ref{fig6},
we plot the enhancement factor, $|\chi_\perp|/\chi_1$, against $\omega$ for
various $r$ and $p$.
Here, in contrast to the absorption (Fig.\ref{fig4}), the enhancement factor
actually increases with the anisotropy as $r$ decreases from unity.
For small $p$, the enhancement peak becomes extremely narrow at $r=0.01$.
However, the enhancement peak must vanish at $r=0$ (because $p_\perp=0$
in this limit).
The separation of the absorption and enhancement peaks is thus even more
pronounced for the perpendicular polarization.
Here we observe that anisotropy has indeed a pronounced effect on the
enhancement peak, as expected.
We conclude that anisotropy can enhance the nonlinearity and its figure of
merit, due to the separation of the absorption peak from the $\chi^{(3)}$
enhancement peak.

We can calculate the optical response of unpolarized light, which is defined
as the average of the results of parallel and perpendicular polarizations.
\begin{eqnarray}
\epsilon_e={1\over 2}(\epsilon_\parallel + \epsilon_\perp), \\
\chi_e={1\over 2}(\chi_\parallel + \chi_\perp).
\end{eqnarray}
The factor 1/2 is due to the fact that unpolarized light has equal
components in the parallel and perpendicular directions, which is valid for
normal incidence and when the anisotropic uniaxial axis is parallel to the
flat surface.
For oblique incidence, we must deal with the $s$ and $p$ polarizations
separately and compute the average. For the $s$ polarization, i.e.,
when the electric field is normal to the plane of incidence, we expect
to obtain the same results as those of normal incidence.
For the $p$ polarization, i.e., when the electric field is parallel to the
plane of incidence, the results will depend on the angle of incidence.

In Fig.\ref{fig7},
the absorption peak Im($\epsilon_e$) for unpolarized light is plotted
against frequency $\omega$ for various values of $r$ and $p$.
In Fig.\ref{fig8},
the enhancement peak $|\chi_e|/\chi_1$ is plotted against frequency $\omega$
for several values of $r$ and $p$.
It should be remarked that the parallel case does not contribute to the
average at $r=0$ because $p_\perp=0$ in this case.
These results may be observed in annealed Au:SiO$_2$ composites
\cite{experiment}.

In order to show the enhancement effect more clearly, we plot in
Fig.\ref{fig9} the figure of merit (FOM), defined by dividing
$\chi_e/\chi_1$ of Fig.\ref{fig8} by Im($\epsilon_e$) of Fig.\ref{fig7}
for the unpolarized case. As is evident from the figures, large FOM
occurs at large $\omega$. However, if we are interested in smaller
$\omega$, say, $\omega/\omega_p<0.5$ because of the quasi-static condition,
the largest FOM of about 120 $\chi_1$ (esu cm) is achieved at $p=0.1$
and $r=0$.

\section*{Discussion and Conclusion}

We have investigated the polarization dependence of optical nonlinearity
enhancement in an anisotropic composite of metal nanocrystals in a
dielectric host.
For the perpendicular polarization, the results are found to be markedly
different from the parallel counterpart.
In contrast to the absorption, the enhancement factor actually increases
with anisotropy.
The separation of the absorption and enhancement peaks is thus even more
pronounced for the perpendicular polarization.

We have computed the spectral density in the anisotropic effective-medium
approximation. The results support our proposal that there can be very
large anisotropy-induced enhancement of nonlinearity.
It is further proposed that such enhancement effect may be realized
experimentally through the application of the electrorheological effect
(e.g., during the annealing process),
with the possibility of achieving even larger optical nonlinearity than that
reported in \cite{experiment}.

It is relevant to note that in this context that the effects discussed in
the present paper are qualitatively different from those of a dispersion
of aligned anisotropic particles (such as sticks) with nonlinear optical
properties. Whereas in the case of aligned sticks the enhancement peak
always coincides with the absorption peak, here the separation of the
two makes the figure of merit much more attractive. Physically, the
difference arises from the different microstructures. In the present case,
the connectedness of the anisotropic microstructure is crucial for the
predicted effects.

Colloidal systems with nonlinear optical particles in the range of 0.1
$\mu$m are good candidates for induced-anisotropy enhancement.
Here the particle size range is determined by considerations that the
particles should be smaller than the optical wavelength, yet large enough
so that the Brownian motion is not strong enough to overwhelm the field
effect (ER effect).

It should be noted that when an intense dc field and an EM field are applied
simultaneously, there can be an enhancement in nonlinearity due to
electrostriction \cite{Neeves}. However, the results of the present work
implies that the induced anisotropy in the microstructure can have an even
more significant enhancement in nonlinearity.

\section*{Acknowledgments}
This work was supported by the Research Grants Council under project
number CUHK 461/95P.
P. Sheng wishes to acknowledge the support of Research Infrastructure
Grant RI93/94.SC09.

\begin{figure}
\caption{Spectral density $m_\parallel(s)$ of parallel polarization of the
anisotropic EMA plotted as a function of $s$, for several values of
$r=0$, 0.25, 0.5, 0.75 and 1.
(a) $p=0.1$, (b) $p=0.3$, (c) $p=0.5$ and (d) $p=0.9$.\label{fig1}}

\vskip 5mm
\caption{Spectral density $m_\perp(s)$ of perpendicular polarization of the
anisotropic EMA plotted as a function of $s$, for several values of
$r=0.01$, 0.25, 0.5, 0.75 and 1.
(a) $p=0.1$, (b) $p=0.3$, (c) $p=0.5$ and (d) $p=0.9$.\label{fig2}}

\vskip 5mm
\caption{The absorption peak Im($\epsilon_\parallel$) plotted against
frequency $\omega$, for several values of $r=0$, 0.25, 0.5, 0.75 and 1.
(a) $p=0.1$, (b) $p=0.3$, (c) $p=0.5$ and (d) $p=0.9$.\label{fig3}}

\vskip 5mm
\caption{The absorption peak Im($\epsilon_\perp$) plotted against
frequency $\omega$, for several values of $r=0.01$, 0.25, 0.5, 0.75 and 1.
(a) $p=0.1$, (b) $p=0.3$, (c) $p=0.5$ and (d) $p=0.9$.\label{fig4}}

\vskip 5mm
\caption{The enhancement peak $|\chi_\parallel|/\chi_1$ plotted against
frequency $\omega$, for several values of $r=0$, 0.25, 0.5, 0.75 and 1.
(a) $p=0.1$, (b) $p=0.3$, (c) $p=0.5$ and (d) $p=0.9$.\label{fig5}}

\vskip 5mm
\caption{The enhancement peak $|\chi_\perp|/\chi_1$ plotted against
frequency $\omega$, for several values of $r=0.01$, 0.25, 0.5, 0.75 and 1.
(a) $p=0.1$, (b) $p=0.3$, (c) $p=0.5$ and (d) $p=0.9$.\label{fig6}}

\vskip 5mm
\caption{For unpolarized light, the absorption peak Im($\epsilon_e$)
plotted against frequency $\omega$, for several values of
$r=0$, 0.25, 0.5, 0.75 and 1.
(a) $p=0.1$, (b) $p=0.3$, (c) $p=0.5$ and (d) $p=0.9$.\label{fig7}}

\vskip 5mm
\caption{For unpolarized light, the enhancement peak $|\chi_e|/\chi_1$
plotted against frequency $\omega$, for several values of
$r=0$, 0.25, 0.5, 0.75 and 1.
(a) $p=0.1$, (b) $p=0.3$, (c) $p=0.5$ and (d) $p=0.9$.\label{fig8}}

\vskip 5mm
\caption{For unpolarized light, the figure of merit plotted against
frequency $\omega$, for several values of $r=0$, 0.25, 0.5, 0.75 and 1.
(a) $p=0.1$, (b) $p=0.3$, (c) $p=0.5$ and (d) $p=0.9$.\label{fig9}}
\end{figure}

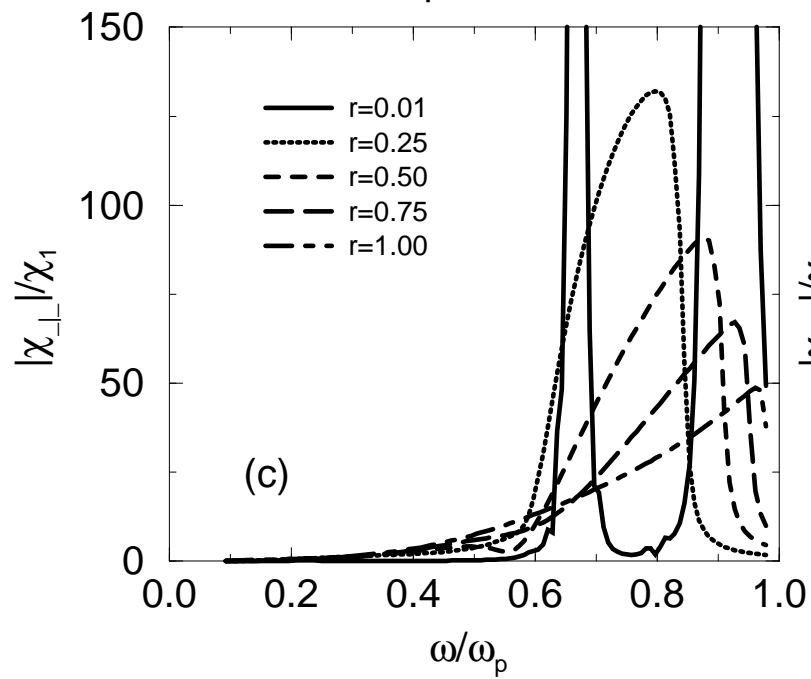
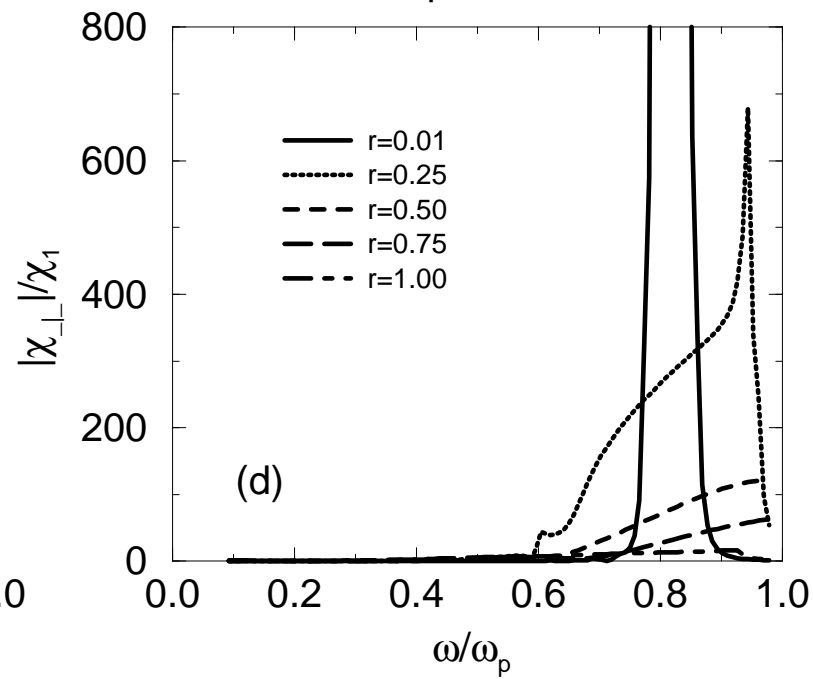
\end{document}